# tubuleTracker: A High-Fidelity Shareware Software to Quantify Angiogenesis Architecture and Maturity

**Authors:** Danish Mahmood[1], Stephanie Buczkowski[1], Sahaj Shah[1], Autumn Anthony[1], Rohini Desetty[1], Carlo R Bartoli[1,2]

[1]Department of Research, Geisinger Medical Center, Danville, PA
[2]Division of Cardiothoracic Surgery, Geisinger Medical Center, Danville, PA

**Corresponding Author:** Carlo Bartoli, MD, PhD
Email: crbartoli@geisinger.edu
Phone: (413) 262-2120

---

## Abstract

**Background:** *In vitro* endothelial cell culture is common to investigate angiogenesis. Histomicrographic images of endothelial cell networks may be quantified with manual analysis. However, this practice is labor intensive and subjective. Shareware software such as ImageJ (National Institutes of Health) is slow and may be inaccurate. Furthermore, as endothelial cell networks become complex, standard architectural metrics may not accurately characterize network maturity. We developed software, *tubuleTracker*, to quantify endothelial cell architecture and maturity quickly and accurately.

**Methods:** Human umbilical vein endothelial cells were grown *in vitro* in extracellular matrix. Phase contrast microscopy acquired cell network images (n=54). Images were evaluated visually with manual analysis by independent reviewers (n=3) as well as with ImageJ and tubuleTracker software. Endothelial cell tubule count, tubule length, node count, tubule area, and vessel circularity were quantified by software. Cell network images were qualitatively scored on a scale of 1-5 (1: most mature network, 5: least mature network) based on the maturity of tubule network formations by trained scientists referred to as "angiogenesis maturity."

**Results:** Statistically significant differences in analysis time (manual ~8 min/image, ImageJ 58±4 s/image, tubuleTracker 6±2 s/image, p<0.0001), tubule count (manual 168±SD, tubuleTracker 92±SD, ImageJ 433±SD, p<0.0001), tubule length (manual 8,117±SD, tubule Tracker 10,894±SD, ImageJ 22,112±SD, p<0.0001), and node count (manual 69±SD, TubuleTracker 77±SD, ImageJ 106±SD, p<0.0001) were observed between the analysis methods. Statistically significant differences in tubule count (p<0.0001), tubule length (p<0.0001), node count (p<0.0001), tubule area (p<0.0001), and circularity of meshes (p<0.0001) were observed between tubule networks with different angiogenesis maturity.

**Conclusions:** tubuleTracker was significantly faster and more accurate than ImageJ and visual analysis of photomicrographs of *in vitro* angiogenesis. Circularity effectively characterized angiogenesis maturity. tubuleTracker is available to the biomedical community as shareware software.

## 1. Introduction

Endothelial cell culture in extracellular matrix is a standard experimental technique to study angiogenesis *in vitro* [1-4]. Histomicrographic images of endothelial cell networks are analyzed to quantify metrics of angiogenesis. Tubule formation is typically photographed with a standard inverted microscope or phase-contrast microscope. Acquired images are analyzed to quantify standard metrics of angiogenesis such as tubule count, tubule length, node count. Manual counting and planimetry of tubules and nodes is relatively simple. However, manual analysis is time consuming and prone to bias [5-7].

Angiogenesis analysis software facilitates faster and potentially more accurate quantification of architectural features of endothelial cell networks in histomicrographs. For example, the ImageJ Angiogenesis Analyzer (National Institute of Health) [8] is frequently used [9-16]. ImageJ shareware software is readily available online, free, and offers many features for a variety of analyses. However, ImageJ has limitations that may affect the accuracy of results. For example, in cultures with a high cell density, ImageJ may improperly identify tubules. Rather than identifying edges that evenly bisect wide monolayers of undifferentiated cells (manually counted as singular tubules in previous studies [2,29,30]), nodes and edges in the network were inappropriately identified [17]. Figure 2d in a study by Gostynska et al shows a similar issue [18]. Several papers which report tubule networks with similar formations of undifferentiated cells do not show the output of ImageJ analysis, so there is reason to believe that inaccurate reports from ImageJ software may be widespread [14,19-23].

As an alternative to ImageJ, pay for service analysis software is available. Wimsas [24] and Angiosys [25] were designed to trace tubule networks from confocal microscopy and tubule count, tubule length, and node count. This software is not readily available, not free, and has not been widely adopted.

Improved software is needed to better quantify angiogenesis in cell culture. The purpose of this study was to develop novel software to more quickly, accurately, and precisely quantify architecture and maturity of endothelial cell networks cultured in vitro. As such, we developed tubuleTracker shareware software to facilitate high-throughput analysis of cell culture experiments for multiple applications across multiple fields. The purpose of this study was to compare the speed and accuracy of tubuleTracker software to ImageJ software and visual evaluation with manual analysis of histomicrographs of in vitro endothelial cell networks.

## 2. Materials and Methods

### 2.1 In Vitro Cell Culture

Third to fifth passage human umbilical vein endothelial cells (HUVECs, Lonza) were cultured in buffered endothelial growth media (EGM-2, Lonza) in a 10-cm culture dish at 37°C with 5% $CO_2$. At 70% confluence, cells were trypsinized and washed with endothelial basal media (EBM, Lonza). HUVECs (8,000 cells/well) were cultured on 50 µL/well Geltrex extracellular matrix (Gibco, ThermoFisher scientific) in a 48 well cell culture plate. After 14 hours of growth, grayscale digital images of endothelial cell networks were randomly acquired with a phase contrast microscope (Zeiss) at 10x. In total, 54 images were acquired and analyzed visually with manual analysis, with ImageJ, and with tubuletracker. Image processing was completed on an AMD Ryzen 7 5700U CPU with 16 GB of DDR4 RAM.

### 2.2 Manual Analysis

Images were evaluated visually with manual analysis by trained laboratory personnel (n=3). A stylus and tablet were used to trace endothelial cell tubules and to identify nodes. Python software (https://anaconda.org/, Anaconda version 3.9) tabulated tubule count, total tubule length, and node count in each image for each viewer.

To characterize endothelial cell angiogenesis maturity, images were ranked on a scale of 1-5. *Mature networks* were characterized by numerous single-cell connections between multiple nodes. Mature networks did not exhibit substantial surface area or common cell alignment. In contrast, *immature networks* were characterized by large sheets of commonly aligned endothelial cells that connected larger but fewer nodes and occupied substantial surface area.

Manual analysis served as control data for comparison to software analysis.

### 2.3 ImageJ Analysis

Fiji ImageJ [26] with Gilles Carpentier's Angiogenesis Analyzer plugin [27]. Histomicrographs were uploaded into the software. Image analysis traced tubules and identified nodes as an overlay on top of the source image. Tubule count, total tubule length, and node count in each image were tabulated in an excel spreadsheet.

### 2.4 tubuleTracker Algorithm and Analysis

The tubuleTracker algorithm converted histomicrographs to a skeleton graph data structure with tubule edges, main nodes, and terminal nodes. Images were denoised with the OpenCV Non-local Means Denoising algorithm. The analog image was converted to a binary with the OpenCV adaptive thresholding algorithm. The c-value of the OpenCV adaptive thresholding algorithm was increased **(Figure 1A-C)** until the amplitude of the lowest frequency in the real-domain of the Fast Fourier Transform (FFT) of the image peaked **(Figure 1D)**, indicating that the tubules were as smooth as possible under adaptive thresholding. However, tubule networks were still sparsely filled, and pixels were not continuously connected. A Gaussian low pass filter was applied to the FFT of the image with a kernel of radius k such that the amplitude of the frequency on the perimeter of the kernel was k/3 **(Figure 1E-H)**. The filter smoothened the tubule network such that OpenCV's otsu thresholding method could be applied to produce a final binary **(Figure 1 I-K)**. The fully connected image was thinned with the SciImage skeletonization function to produce a 1-dimensional medial skeleton of the network surface **(Figure 1 K, L)**. Main nodes in red were identified as pixels in the skeleton with three or more neighbors. Edges in orange were identified as paths on the skeleton between the main nodes. Terminating nodes in green were identified as pixels on the skeleton with only one neighbor. Edges in green were identified as paths on the skeleton between a main node and terminating node. Pseudo-nodes in gray were not counted after removing small terminating node artifacts from the image. Tubule count, total tubule length, node count, and tubule area in each image were tabulated in an excel spreadsheets.

**2.5 Endothelial Cell Network Circularity: A Novel Metric of Angiogenesis maturity**

For each image, the best fit ellipse for non-populated regions were defined with an OpenCV regression function. The inverse of the difference between the area of the best fit ellipse and the actual non-populated area was quantified as network "circularity" to be compared with standard metrics in relationship qualitative assessment of "angiogenesis maturity."

**2.6 Statistics**

Statistics were performed with (python, 3.9, Anaconda). One-way, non-repeated measures ANOVA with Tukey post-hoc test was used to compare the tubule count, tubule length, and node count quantified by manual analysis, ImageJ analysis, and tubuleTracker analysis. One-way, non-repeated measures ANOVA with Tukey post-hoc test was used to compare the tubule count, tubule length, node count, tubule area and circularity of histomicrographs with varying angiogenesis maturity. Unpaired two-way Student's t-tests were used to compare speed of analysis between softwares and total tubule area. A $p<0.05$ was considered statistically significant.

## 3. Results

### 3.1 tubuleTracker is faster at processing images than imageJ and manual visual analysis

Figure 2 shows that the mean processing time of tubuleTracker for each image was 6±2 s/image. The mean processing time for ImageJ for each image was 58±4 s/image. The mean time taken by trained investigators for manual analysis of the images was 8 min/image. Tubule Tracker was significantly faster ($p<0.001$, $n=54$) at processing images than imageJ.

### 3.2 Tubule Count, Tubule Length, and Node Count measured by tubuleTracker closer to manual visual analysis than imageJ

Figure 4 shows that statistically significant differences in tubule count (manual 168±SD, tubuleTracker 92±SD, ImageJ 433±SD, $p<0.0001$), tubule length (manual 8,117±SD, tubule Tracker 10,894±SD, ImageJ 22,112±SD, $p<0.0001$), and node count (manual 69±SD, TubuleTracker 77±SD, ImageJ 106±SD, $p<0.0001$) were observed between the analysis methods. Statistically significant differences in tubule count ($p<0.0001$), tubule length ($p<0.0001$), node count ($p<0.0001$), tubule area ($p<0.0001$), and circularity of meshes ($p<0.0001$) were observed between tubule networks with different angiogenesis maturity.

### 3.3 Relationship between software metrics and gold-standard manual classification

Statistically significant differences in tubule count ($p<0.0001$), tubule length ($p<0.0001$), node count ($p<0.0001$), tubule area ($p<0.0001$), and circularity of meshes ($p<0.0001$) were observed between tubule networks with different angiogenesis maturity.

## 4. Discussion

We developed novel software to analyze histomicrographs and quantify endothelial cell network architecture. Specifically, we observed that tubuleTracker software was (1) statistically faster than ImageJ software and visual analysis, (2) statistically more accurate than ImageJ at quantifying tubule count, tubule length, node count (3) treats wide monolayers of cells as individual tubules. Furthermore, we observed that tubule area and mesh circularity were more strongly correlated with angiogenesis maturity while tubule count, tubule length, and node count were not linearly correlated.

### 4.1 tubuleTracker is Faster and More Accurate than ImageJ

tubuleTracker software analyzes images quickly. As in figure 2, approximately 6.1 seconds were needed to process each image, which was 9.5 times faster than ImageJ. Further reduced image processing times are likely easily achieved. ImageJ is a production software compiled in JAVA. In contrast, tubuleTracker is a prototype python script that is interpreted and contains excess bulk from package components which are not used. Complied languages are inherently faster than interpreted languages. As a result, it is likely that tubuleTracker may be even faster if compiled in production code. Furthermore, the utility of python script exceeds ImageJ software because it is easily scriptable for batch analysis of entire experiments of data in a directory.

In figure 4, it is observed that the difference in tubule count reported by tubuleTracker and manual analysis was significantly higher than that reported by ImageJ and manual analysis. The difference in tubule length reported by tubuleTracker and manual analysis was significantly higher than the difference reported by ImageJ and manual analysis. Node count reported by tubuleTracker was not significantly different from visual analysis whereas node count reported by ImageJ was significantly higher than that reported by visual analysis.

In technical terms, ImageJ software improperly assesses tubule networks because of a limitation in converting confocal microscopy images to a binary. Binaries are generated based on the contrast between dark and light pixels, so cell nuclei and their membrane extensions become "on" pixels while the background becomes "off" pixels. The binary is converted to a skeleton, and nodes are identified as pixels with three or more neighbors and edges are identified as a set of pixels in the skeleton connecting nodes [27]. Hence, a wide sheath of undifferentiated cells may result in a sparsely filled binary which becomes a scattered skeleton that is then traced.

## 4.2 Network Circularity of vessel cross sections and tubule area correlate more strongly with angiogenesis than tubule count, tubule length, node count

Traditional metrics of angiogenesis reported by software may be insufficient to quantify endothelial cell network maturity. Following plating onto Matrigel, endothelial cells form cellular extrusions which roll up into networks of capillary-like aggregates or cords [3]. Arnaoutova et al reports the different ways tubule networks form as a function of plating density, basement membrane extract, and time [28]. In the paper, Figure 2 indicates that when cell density is low (4800 cells/cm$^2$) the node/hub count decreases as cells incubate over time (1 hours to 24 hours). Intuitively, cells are initially separate entities which form extensions to one another, migrate towards one another, and roll up into tubules, so fewer "hubs" exist. Figure 3 indicates that higher cell plating density increases the perceived thickness of tubules (i.e. tubule area) after incubation over a specified period of time. Figure 5 indicates that as greater volumes of basement membrane extract are used (i.e. Geltrex in this paper), cells are stimulated to differentiate and roll up into tubules instead of growing uniformly on the dish as undifferentiated cell clusters. The number of enclosed regions/meshes (empty space surrounded by cells) increases as cells differentiate. This can also be seen in Figure 6 where angiogenesis stimulating factors in the basement membrane extract are modulated.

Hence, we proposed that tubule area and circularity (defined previously as the best fit ellipse of vacant regions in the network) may be more important than standard reported metrics (tubule count, length, and node count) for tubule networks with a high density of cells. To test this, trained scientists were asked to qualitatively rank the angiogenesis maturity of the network (i.e. differentiation level of cells) on a scale of 1-5, where 1 is the most differentiated and 5 is the least differentiated.

Figure 3 shows that cell tubule count, tubule length, and node count increased with higher visual ranking of endothelial cell angiogenesis maturity from 1 and 3 but decreased for maturity rankings of 4 through 5. Hence, these metrics alone are not sufficient to assess angiogenesis maturity. Tubule area and vessel circularity increased for angiogenesis maturity rankings of 1 through 5. We found vessel circularity and tubule area to continue increasing with more mature endothelial cell networks.

## 5. Conclusions

*In vitro* endothelial cell culture is a common experimental tool to investigate angiogenesis. Visual analysis of cell culture histomicrographs is slow and cumbersome. Current shareware software slowly and inaccurately quantifies angiogenesis in cell culture. We developed novel software that quickly and accurately quantified standard metrics of *in vitro* angiogenesis. tubuleTracker is available as shareware software to improve analysis of *in vitro* angiogenesis assays and facilitate high-throughput angiogenesis experimentation for multiple applications across multiple fields.

## Acknowledgements

We acknowledge Patrick Hill for administrative support.

**Funding:** Funding was provided in part by a grant from the Children's Hospital of Philadelphia Cardiac Center and Geisinger Medical Center.

**Disclosures:** The authors declare no disclosures or conflict(s) of interest.

**Figure Legends**

**Figure 1: A)** raw image **B)** adaptive thresholding with c-value 1 **C)** adaptive thresholding with c-value 5 **D)** magnitude of lowest frequency of FFT with increasing c-values. Red line indicated c-value used for downstream processing **E)** FFT Magnitude spectrum of adaptive thresholding input with c-value 4 **F)** adaptive thresholding input with c-value 4 **G)**Gaussian Kernelled FFT Magnitude spectrum of adaptive thresholding input with c-value 4 **H)** Low pass filter output (inverse FFT of Gaussian Kernelling) **I)** Otsu binary thresholding output **J)** Skeletonization of binary thresh-holding **K)**Traced skeleton with nodes and edges. **L)** Node count decreasing with each iteration of small branch removal

**Figure 2:** Image processing time of manual analysis (red line), ImageJ, and tubuleTracker.

**Figure 3:** Tubule Count, Tubule Length, Node Count, Tubule Area, Angiogenesis Circularity measured by tubuleTracker of n=54 images compared to manually classified maturity on a scale of 1-5.

**Figure 4:** Tubule Count, Tubule Length, and Node Count measured by Manual Analysis, Tubule Tracker, and ImageJ.

**Figures**

**Figure 1**

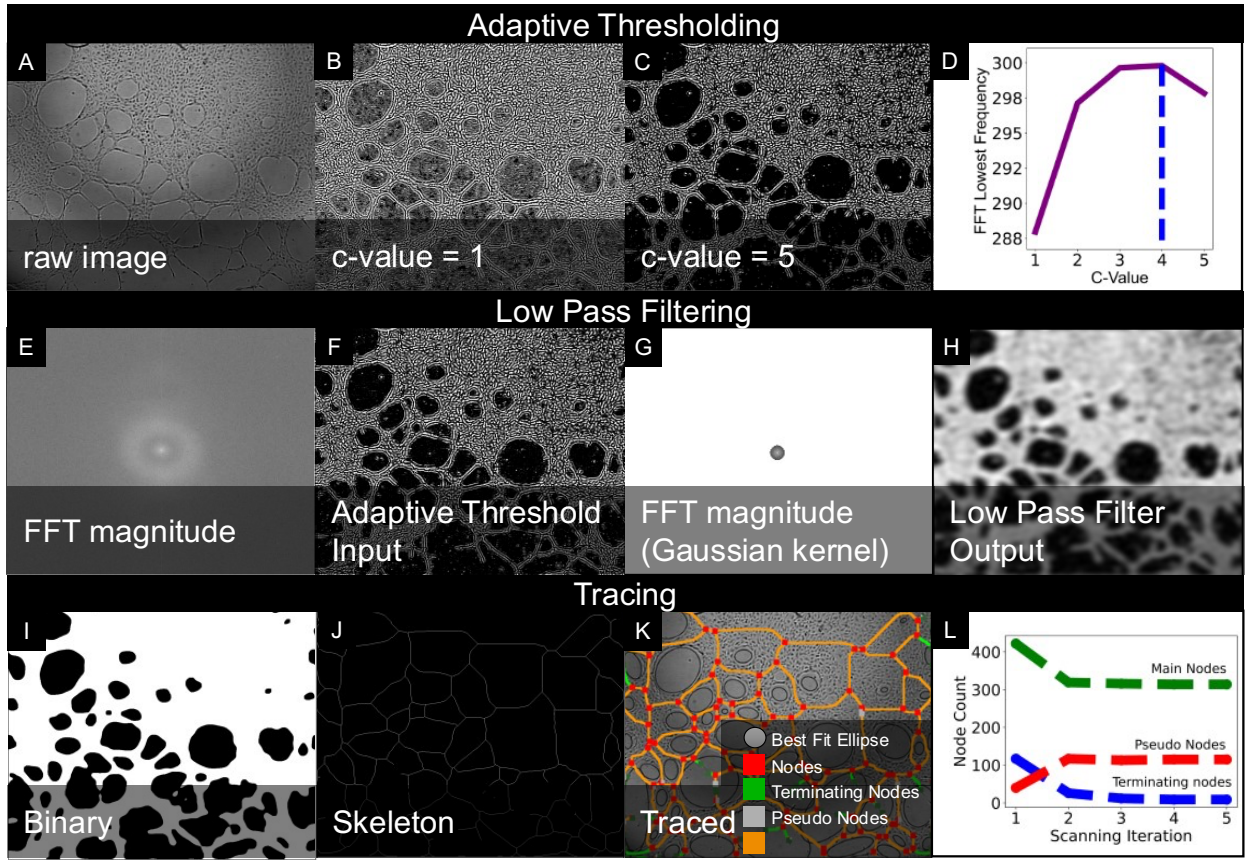

**Figure 2**

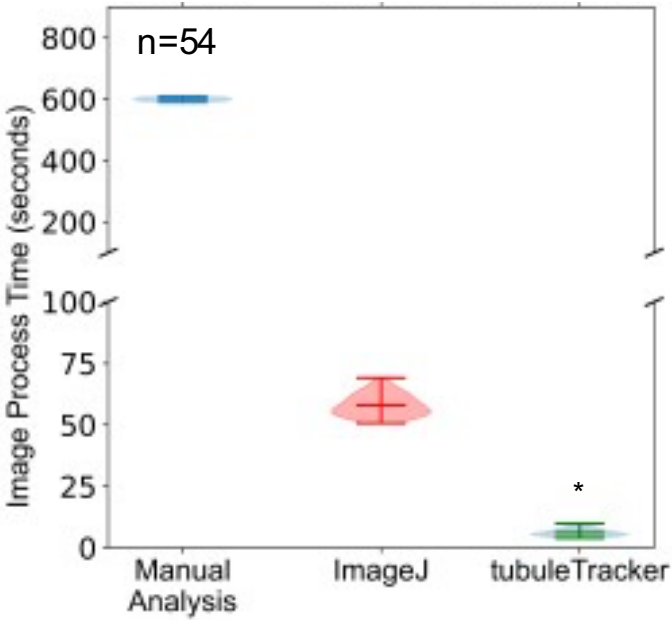

* ImageJ vs tubuleTracker p<0.0001

**Figure 3**

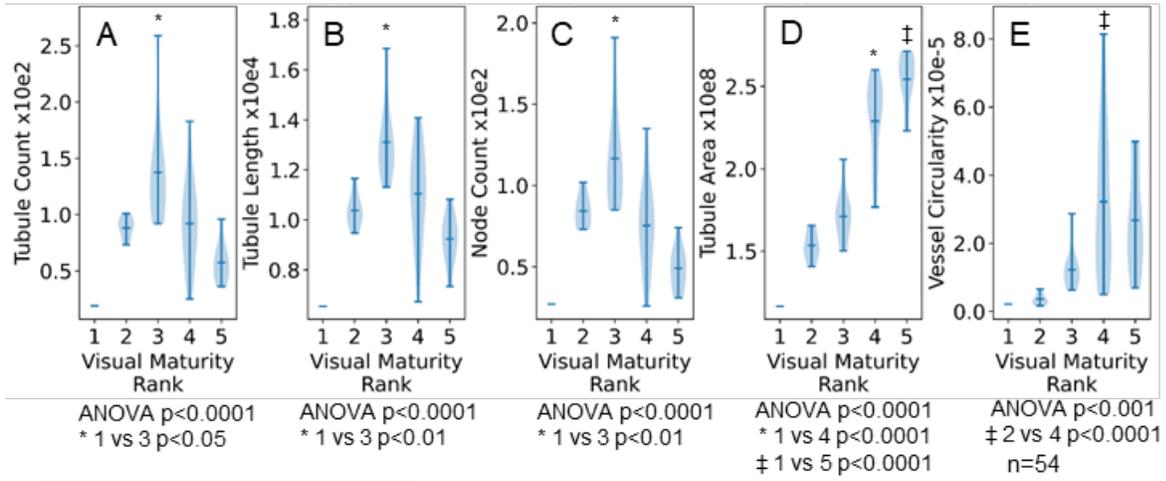

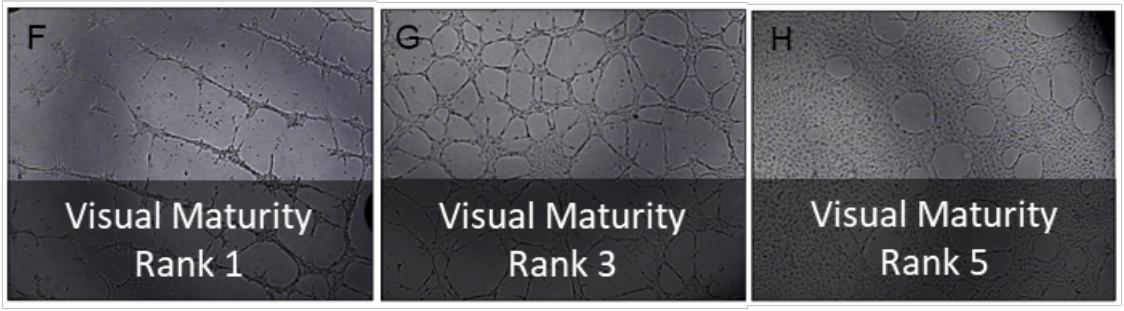

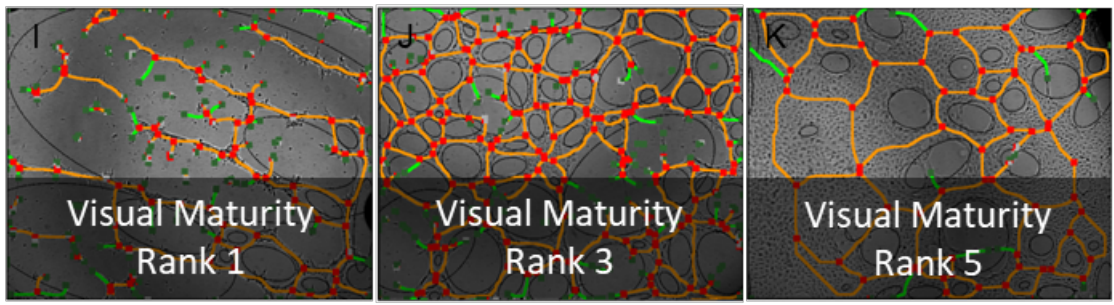

**Figure 4**

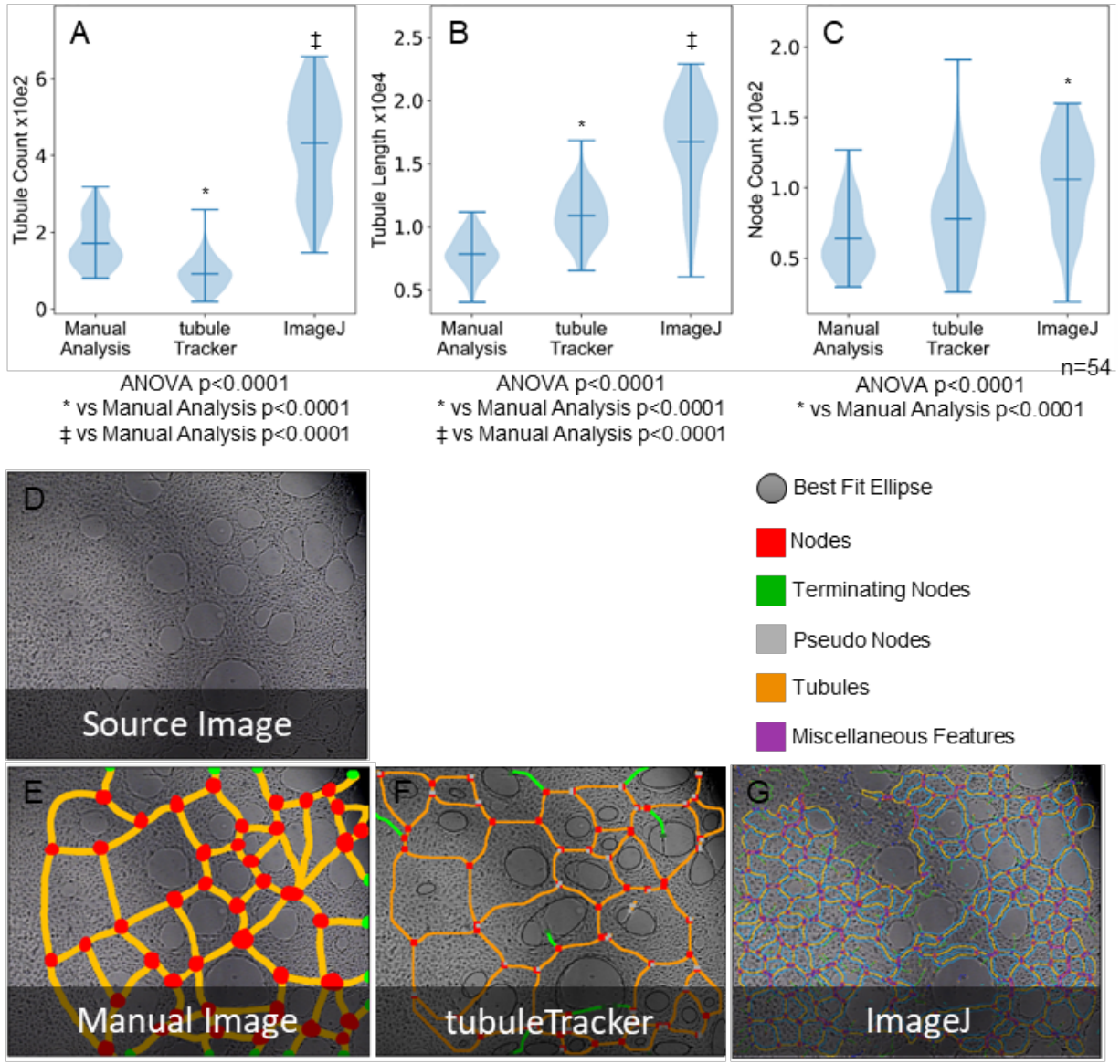